{Commentary}

# Recommendations to enhance rigor and reproducibility in biomedical research

Jaqueline J. Brito[1,*], Jun Li[2], Jason H. Moore[3], Casey S. Greene[4,5], Nicole A. Nogoy[6$], Lana X. Garmire[2$], Serghei Mangul[1,$*]

[1] Department of Clinical Pharmacy, School of Pharmacy, University of Southern California, 1985 Zonal Avenue Los Angeles, CA 90089, USA

[2] Department of Computational Medicine & Bioinformatics, Medical School, University of Michigan, 1301 Catherine St. Ann Arbor, MI 48109, USA

[3] Department of Biostatistics, Epidemiology, and Informatics, Institute for Biomedical Informatics, University of Pennsylvania, 3700 Hamilton Walk, Philadelphia, PA 19104, USA

[4] Department of Systems Pharmacology and Translational Therapeutics, Perelman School of Medicine, University of Pennsylvania, 3400 Civic Center Blvd, Philadelphia, PA 19104, USA

[5] Childhood Cancer Data Lab, Alex's Lemonade Stand, Philadelphia, PA 19102, USA

[6] GigaScience, 26/F, Kings Wing Plaza 2, 1 On Kwan Street, Shek Mun, N.T, Hong Kong

$ - These authors contributed equally to the paper

*Correspondence: britoj@usc.edu; serghei.mangul@gmail.com

**Abstract**

Computational methods have reshaped the landscape of modern biology. While the biomedical community is increasingly dependent on computational tools, the mechanisms ensuring open data, open software, and reproducibility are variably enforced by academic institutions, funders, and publishers. Publications may present academic software for which essential materials are or become unavailable, such as source code and documentation. Publications that lack such information compromise the role of peer review in evaluating technical strength and scientific contribution. Incomplete ancillary information for an academic software package may bias or limit any subsequent work produced with the tool. We provide eight recommendations across four different domains to improve reproducibility, transparency, and rigor in computational biology—precisely the main values which should be emphasized in life science curricula. Our recommendations for improving software availability, usability, and archival stability aim to foster a sustainable data science ecosystem in biomedicine and life science research.

**Main text**

Biomedical informatics is increasingly becoming essential to development of practices that promote open data, open software, and reproducible research in the scientific community. Computational reproduction of previously published results is enabled when scientists publicly release all research resources, from raw data to installable packages and source code, in a discoverable and archivally stable manner. Publications lacking data or source code sharing undermine scientific rigor, transparency, and reproducibility[1]. Platforms already exist that support public release of scientific materials, but the current lack of strict enforcement by journals, academic institutions, and funding agencies has resulted in a loss of essential data and source code for many published studies.

An astonishing number of bioinformatics and computational biology software tools are designed each year to accommodate increasingly bigger, complex, and specialized biomedical datasets[2]. Many of those software tools have limited installability, are closed-source, or are hosted on Uniform Resource Locators (URLs) with undetermined archiving protocols[3]. Lack of access to the source code of a software package undermines the auditing of methods and results and ultimately harms the transparency of research. Prior studies[4] have addressed issues of computational reproducibility, including the need to automatize all data manipulation tasks and version control of code. We expand upon existing dialogue and emphasize reproducible research as computational training, journal policies, and financial support. We identify and discuss eight key recommendations across four different domains (**Figure 1**) to tackle the pressing need for scientists to improve software availability, usability, and archival stability in computational biology. By following a set of best practices[5], scientists can promote rigor and reproducibility, ultimately cultivating a sustainable, thriving research community.

## 1. Teaching computational skills to produce reproducible research

**Increase computational training opportunities targeted at reproducibility.** Biomedical researchers who use computational tools must acquire specific computational skills in order to successfully apply the techniques to a large amount of data. Undergraduate students who lack formal computational training can be taught the skills required to promote reproducibility via specialized courses. In addition to rigorous class training, advanced undergraduate and graduate students, postdoctoral scholars, clinical fellows, and faculty may benefit from short-term intensive workshops. Several institutions, including the University of California, Los Angeles, have successfully hosted workshop-based programs for over five years and serve as valuable resources for pedagogy and curriculum development[6]. Effective workshops for training researchers to use computational tools include curated, hands-on training experiences for implementing analysis tools, such as interactive cloud-based notebook technologies. Since 1998, Software Carpentry (https://software-carpentry.org/) has been holding volunteer-based training courses for researchers who wish to master the computational skills required to keep up with the demands of data- and computational-intensive research. Today's biological researcher must learn to use the command line in order to run analyses in open-source software packages. ,

Comprehensive computational training programs are ideal platforms for training future life science and biomedical researchers in techniques that support reproducibility **(Figure 1a)**.

## 2. Development and distribution of data and software

**2.1. Make all data and metadata open and discoverable.** Open source code depends on the availability of open and shareable data, and access to the data used to produce important research results is key for auditing the rigor of published studies. Open access to datasets is imperative to building a thriving and sustainable scientific community where all researchers can access and analyze existing data. In practice, omics data of patients often cannot be publicly shared due to patient privacy and/or user agreement standards[7]. While not all data are freely and publicly available, many studies provide controlled data access where researchers can sign a user agreement to access the raw data once their scientific rationale is approved. In general, the global data sharing climate has shifted towards a positive direction; even in cases where raw data are not accessible by the public, summary data are often available.

Truly open data sharing supports the reproducibility and robustness of science because it enables others to reuse data on larger-scale analyses. In addition, secondary analysis is an economically sustainable approach that can be adopted by scientists in countries or at institutions with limited computational resources[8]. Ideally, data should also be discoverable via centralized repositories, such as Sequence Read Archive (SRA) and Gene Expression Omnibus (GEO), and annotated with descriptive metadata to enhance data reuse **(Figure 1b)**. When data are shared on centralized repositories in interoperable formats, other researchers can examine and re-analyze the data, challenge existing interpretations, and test new theories. Data sharing corresponds to the true spirit of science, where each new discovery is built upon previous work and ultimately allows us to "stand on the shoulders of giants". Many important scientific discoveries have been solely based on shared data (e.g., economics, meteorology, and physics). Reusing data further emphasizes the quality and importance of generated data and contributes to the impact of the original, data-generating research.

**2.2. Build and use open-source software.** Software provides a foundation for scientific reproducibility—the ability to replicate published findings by running the same computational tool on data generated by the study[4]. Open-source academic software are advantageous to the scientific community, because closed-source proprietary software restricts the reproducibility of biomedical research. First, lack of access to the source code limits other researchers' ability to audit results and reviewers' ability to test the reproducibility prior to publication. Second, license restrictions may prohibit the creation of new functionalities that could be released on modified versions of existing tools. Not every laboratory or researcher can afford the cost of acquiring and maintaining proprietary software licenses. Reviewers may lack access to proprietary software and be unable to fully test the reproducibility of results. Widespread adoption of standard open-source licenses for data and software tools can enhance the rigor and impact of research by allowing any researcher and reviewer to reproduce published studies.

Publicly releasing the source code does not guarantee the computational reproducibility of biomedical research. Software must be well documented with user manuals and installable in a user-friendly manner. Code used in a published analysis should be hosted on an archivally stable platform such as Software Heritage Archive (https://archive.softwareheritage.org/) or Zenodo (https://zenodo.org/) (**Figure 1c**). Currently, over one-fourth of computational software resources cannot be accessed through the URLs provided in the original publication, suggesting that the repositories are poorly maintained[3]. Additionally, many bioinformatics tools are too difficult, or even impossible, for a new user to install[3]. Use of Open Source Initiative license models (https://opensource.org/licenses) allows users to easily use and adapt tools, increasing the sustainability of the biomedical research community. Hosting software tools on package managers allows users to easily install software with more straightforward commands and automatically acquire resolutions for software dependencies. Examples of package managers are Conda and Bioconda **(Table 1).**

**2.3. Leverage platforms that enhance reproducibility.** In addition to software and datasets, computational biology researchers commonly produce resources such as experiment protocols, workflows, and annotations. Storing and sharing these resources on a stable platform allows other researchers to cite the materials, which would increase the reproducibility of a paper and the visibility of previously developed methods. The inclusion of citable digital object identifiers (DOIs) also facilitates the discovery of reusable resources as they provide long-term access to published resources. Several innovative platforms designed to promote reproducibility have recently emerged (**Figure 1d**).

## 3. Implementation of reproducible research

**3.1. Make tools and workflows reproducible.** Virtual machines (VMs) and containers can be used to facilitate the reproducibility of open-source software tools. VMs are software pieces that are capable of encapsulating entire operating systems, libraries, codes, and data. Reproducibility can be enhanced with workflow-specific platforms, such as Galaxy and Tensorflow (for machine learning), and workflow standards, such as CWL (Common Workflow Language) **(Table 1)**. Various platforms and tools are now available that support reproducible research and are already commonly used by life science and biomedical researchers (**Table 1**). Given the many different tools and platforms available, a research lab should define their own standards on a suite of tools and platforms that support their research practices (**Figure 1e**).

**Table 1: Examples of tools and platforms to share reproducible resources.**

| Platform & Type | Use |
|---|---|
| **Reproducible and open methods** | **Protocols.io (**RRID:SCR_010490**)** is an open-source protocol repository, where researchers can manage, share, tweak, |
| | optimize, and adopt existing methods even after a scientist has left a lab. |
| **RRIDs** | **Scicrunch.org** is a platform for curating research resources that enables the user to discover, access, view, and use research objects. Users can register any research object, such as tools, antibodies, and animal models. In turn, these objects are issued a Research Resource ID (RRID), which should be cited in the manuscript. The RRID allows other users to easily locate and access the resources. |
| **Annotations** | **Hypothes.is** (RRID:SCR_000430) is an open-source annotation tool that allows any researcher to annotate any resource on the web, for personal use or as part of conversations available to private groups or the general public. |
| **Virtual Machines & Containers** | Containers such as **Docker** ([www.docker.com](www.docker.com)), and **Singularity** ([singularity.lbl.gov](singularity.lbl.gov)) are lightweight solutions compared to VMs as they do not encapsulate the operating system; rather, they rely on the host kernel to run required functions. Both VMs and containers are shared via image files and can be included as supplementary material at certain journals or stored in Zenodo |
| | (https://zenodo.org/, RRID:SCR_004129), Figshare (https://figshare.com/, RRID:SCR_004328), or other general-purpose archival repositories. |
| **Reproducible workflows** | **Galaxy** (https://galaxyproject.org/, RRID:SCR_006281) is a computational platform which allows users to share workflows, histories, and wrapped tools in an easy-to-use and open-source interface that even people without coding experience can use. **Common workflow language (CWL)** (https://www.commonwl.org, RRID:SCR_015528) is an open standard used to describe workflows and tools to make them portable and interoperable across different environments (e.g., cloud, cluster, or high-performance computing). **Tensorflow** (https://www.tensorflow.org, RRID:SCR_016345) is an open source end-to-end machine learning platform with broad use (e.g., data, library and neural networks). Tensorflow provides workflows to develop and train models using many other programming languages. **Snakemake** (https://snakemake.readthedocs.io/en/stable/, RRID:SCR_003475) is a tool to create reproducible and scalable data-analyses workflows, with a language based on python. Snakemake makes it easier to execute data analyses on different |
| | environments without modification on the workflow definition. |
| **Package managers** | **Conda** (https://conda.io/, RRID:SCR_018317) is a powerful open source package and management system that can quickly install, run, and update packages and their dependencies. **Bioconda** (https://bioconda.github.io/, RRID:SCR_018316) leverages Conda and is a community project and package manager dedicated to computational tools used by life science and biomedical researchers. |
| **Reproducible documents & Figures** | **Jupyter Notebook** (https://jupyter.org/) allows for the creation of sharing of live code, equations, visualizations, and narrative text. The application supports over 40 different programming languages and can be used to leverage big data. **MyBinder** (https://mybinder.org/, RRID:SCR_016437) is an application that collects and 'binds' interactive jupyter notebooks into a Binder repository and can also create a Docker image of the collection. **Stencila** (https://stenci.la/) is an open source framework for executable documents and living figures (using R scripts). It supports commonly used environments and tools, such as Jupyter Notebook, RMarkdown, Python, and SQL. |

**3.2. Implement living and reproducible figures and papers.** Archiving open data and code is an important fundamental step toward transparency; however, over the last 5 years, it has been possible to break away from the static presentation of results and produce dynamic, or "living," figures (**Figure 1f**). Dynamic figures allow a reader to alter parameters of an analysis as the code is actively running—an iterative process where a data visualization can evolve in real time as new data is added. One such example is Stenci.la, a platform that supports executable documents, living figures, and Jupyter Notebooks (**Table 1**).

## 4. Incentivizing reproducible research

**4.1. Enforce reproducibility upon the peer-reviewing process.** Journals have various publishing standards. Stakeholders from academia and industry have defined a set of principles stating that research data should be Findable, Accessible, Interoperable, and Reusable (FAIR)[9] (**Figure 1g**). Researchers may elect to publish in journals that encourage best practices (e.g., adopting the FAIR principles[9]) that aim to increase the impact of their work. To ensure reproducibility, many journals now require that biomedical data generated by a published study be shared when the paper is released. For instance, *GigaScience* (gigasciencejournal.com) has been promoting reproducibility of analyses since 2012 (in addition to publishing open access) by mandating open data arrangements that follow the FAIR principles and mandates availability of source code with an OSI approved license. During peer review, *GigaScience* makes all supporting data and code available for reviewers, and editors ask reviewers to test provided materials for reproducibility. Authors can aid this task by including VMs, containers, Jupyter Notebooks, or packaged workflows (as opposed to static versions of these resources). *Biostatistics* has begun issuing badges for articles with validated data and code sharing**.** In 2018, eLife published a demonstration of a dynamic and code-based reproducible peer-reviewed paper, using the Stencila platform and Binder (**Table 1**). This approach enables data and analysis to be fully reproducible by the reader and challenges the traditional static representation of results using PDF or HTML formats.

**4.2. Create earmarked funds and reporting requirements to support reusable resources**

Successfully implementing and widely distributing software tools developed in academia involves unique challenges when compared to doing so in industry. In academia, software tools are developed by small groups comprised of graduate or postdoctoral scholars. These groups have fairly fast turn-over rates of 2-5 years and are less likely to be professionally trained in software production standards. In industry, software development groups are comprised of holistic teams of specialists capable of supporting long-term software maintenance. In order to enhance the quality and reuse of open software, academic groups should hire professionally trained software engineers to partner with students and postdocs. Clearly hiring industry software developers represents a burden on academic teams; funding agencies need clear mechanisms of acknowledging and incentivizing funding earmarked for critical bioinformatics infrastructure (**Figure 1h**). In addition, funders should recognize the rigor of software development, rather than just considering 'novelty'-based conventional criteria of research. The availability of well-resourced grant mechanisms to convert minimum viable products produced by trainees into reliable software could enhance the impact of research-grade software on the community. With the growing number of biomedical datasets open for reuse in the public domain, it is inspiring to see the encouragement and acknowledgment of data reuse and secondary analysis with the Research Parasite Awards[10]. The annual Parasite Awards recognizes the exceptional contributions for rigorous secondary analysis of data with recognition of the top-performing junior parasite and senior parasite. More initiatives, such as this, are needed for promoting software reuse.

**Conclusions**

We outlined eight key recommendations across four different domains to improve the rigor of biomedical studies and foster reproducibility in computational biology. The infrastructure required to systematically adopt best practices for reproducibility of biomedical research is largely in place; the remaining challenge is that incentives are not currently aligned to support good practices. Instead, current efforts rely on individual researchers electing to follow the best practices, often at their own time and expense. We believe it is time for a fundamental cultural shift in the scientific community: rigor and reproducibility should become primary concerns in the criteria and decision-making process of designing studies, funding research, and writing and publishing results. Successful systematic adoption of best practices will require the buy-in of multiple stakeholders in the scientific communities, from publishers, academic institutions, funding agencies, and stakeholders. Such commitment would increase the lifetime and scientific value of published research as resources naturally become reusable, testable, and discoverable.

Community-wide adoption of best practices for reproducibility is critical to realizing the full potential of fast-paced, collaborative analyses of large datasets in the biomedical and life sciences. The platforms listed in this paper are provided for illustration. Given this is a fast-moving area, some of our recommendations are likely to be outdated within a short period and other short-lived. We acknowledge that new platforms may appear soon (https://github.com/Mangul-Lab-USC/enhancing_reproducibility).

**Abbreviations**

CWL, Common Workflow Language; DOI, Digital Object identifier; FAIR, Findable, Accessible, Interoperable, and Reusable; GEO, Gene Expression Omnibus; OSI, Open Source Initiative; RRID, Research Resource ID; SRA, Sequence Read Archive; URLs, Uniform Resource Locators; VM, Virtual Machine.

**Competing interests**

The authors declare that they have no competing interests.

**Funding**

CSG was supported by grants from the NIH (R01HG010067 and R01CA237170), the Gordon and Betty Moore Foundation (GBMF4552), the Chan Zuckerberg Initiative Donor Advised Fund of the Silicon Valley Community Foundation (2018-182718), and Alex's Lemonade Stand Foundation (CCDL). LXG is supported by grants K01ES025434 awarded by NIEHS through funds provided by the trans-NIH Big Data to Knowledge (BD2K) initiative

(http://datascience.nih.gov/bd2k), R01 LM012373 and R01 LM012907 awarded by NLM, and R01 HD084633 awarded by NICHD. The funding bodies had no role in the design of the study and collection, analysis, and interpretation of data and in writing the manuscript.

**Authors' contributions**

| Author | Contribution |
| --- | --- |
| JJB | Writing - Original Draft Preparation, review and editing. |
| JL | Writing - review and editing. |
| CSG | Writing - review and editing. |
| JM | Writing - review and editing. |
| NN | Conceptualization and structure of the manuscript; Writing - review and editing; Visualization - creation of Figures and Table. |
| LG | Conceptualizing the project; Writing - review and editing. |
| SM | Conceptualization and structure of the manuscript; Writing - review and editing. |

**Acknowledgments**

We thank Dr. Lana Martin for helping with the design of Figure 1 and constructive discussions and comments on the manuscript. A dynamic version of this paper reflects new platforms

[https://github.com/Mangul-Lab-USC/enhancing_reproducibility](https://github.com/Mangul-Lab-USC/enhancing_reproducibility), as an example to command scientific rigor and reproducibility ourselves. This dynamic version was compiled in markdown and includes an extended list of references. We encourage others to contribute to our repository.

**Authors' Information**

NAN is an Editor at *GigaScience* and is an open science advocate with over 8 years experience in publishing reproducible research.

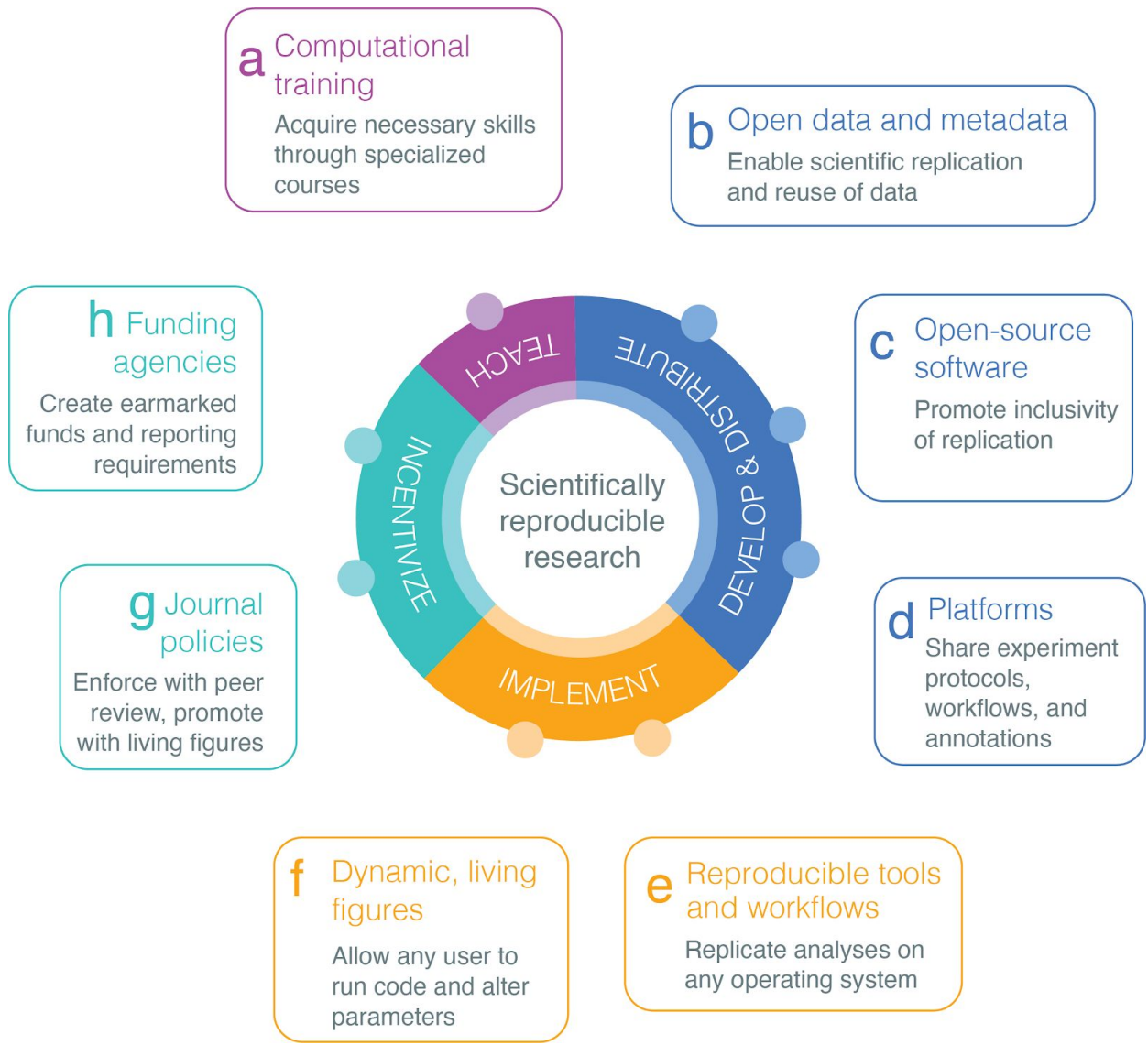

**Figure 1.** Recommendations to improve reproducibility and rigor of biomedical research organized across the four domains: Teaching computational skills to produce reproducible research ("Teach"); Development and distribution of data and software ("Develop and Distribute"); Implementation of reproducible research ("Implement"); and Incentivizing reproducible research ("Incentivize").